\documentclass[a4paper,11pt,twoside]{article}
\usepackage[paperheight=29.7cm,paperwidth=21cm,outer=2.25cm,inner=2.25cm,top=2cm,bottom=2cm,headheight=13.6pt,includehead,includefoot]{geometry}
\usepackage{silence}
\newcommand{\textlineskip}{\baselineskip=13pt}

\providecommand{\keywords}[1]{\noindent\textbf{Keywords:} #1}

\def\fnt#1#2{\footnotetext{\kern-.3em%
          {$^{\mbox{\scriptsize #1}}$}{#2}}}

\usepackage{fancyhdr}

\fancypagestyle{pagestyle}{%
  \fancyhf{}
  \fancyhead[OR]{E. Tolotti, E. Blanzieri, D. Pastorello}
  \fancyhead[EL]{A weighted quantum ensemble of homogeneous quantum classifiers}
  \fancyfoot[C]{\thepage}%
}

\fancypagestyle{firstpagestyle}{%
  \fancyhf{}
  
  \fancyfoot[C]{\thepage}
}

\renewcommand{\thefootnote}{\fnsymbol{footnote}}  

\usepackage{amsmath}
\usepackage{amssymb}
\usepackage[ruled, linesnumbered]{algorithm2e}
\usepackage[font={footnotesize}]{caption}

\usepackage{bm}

\usepackage{multirow}
\usepackage[figuresright]{rotating}

\usepackage[hyphens]{url}
\usepackage[hyperfootnotes=false]{hyperref}
\usepackage{xcolor}
\hypersetup{
  colorlinks,
  linkcolor={black},
  citecolor={black},
  urlcolor={black}
}
\usepackage[noabbrev,capitalise]{cleveref}

\usepackage{csquotes}
\usepackage[
backend=biber,
style=ieee,
sorting=none,
url=false,
eprint=false
]{biblatex}

\usepackage{soul}

\usepackage{etoolbox}
\usepackage{accents}
\robustify{\underaccent}

\usepackage{braket}

\usepackage{tikz}
\usetikzlibrary{quantikz2}

\usepackage[caption=false]{subfig}

\usepackage{todonotes}

\begin{document}

\newcommand{\xtest}{x}
\newcommand{\ytest}{y}
\newcommand{\xtrain}{x}
\newcommand{\ytrain}{y}

\normalsize\textlineskip
\thispagestyle{firstpagestyle}
\setcounter{page}{1}

\centerline{\Large\bf
A weighted quantum ensemble of homogeneous quantum classifiers}

\vspace{25pt}

\begin{center}
    \textbf{Emiliano Tolotti$^{\star \|}$}, \hspace{1pt}
    \textbf{Enrico Blanzieri$^{\star \dagger}$}, \hspace{1pt}
    \textbf{Davide Pastorello$^{\ddagger \dagger}$}

    \vspace*{10pt}

    $^\star$ Department of Information Engineering and Computer Science\\ University of Trento \\ 
    $ $ via Sommarive 9, 38123 Povo, Trento, Italy

\vspace*{10pt}

    $^\ddagger$ Department of Mathematics\\ Alma Mater Studiorum - Università di Bologna \\ 
    $ $ Piazza di Porta San Donato 5, 40126 Bologna, Italy

    \vspace*{10pt}

    $^\dagger$ Trento Institute for Fundamental Physics and Applications \\ 
    $ $ via Sommarive 14, 38123 Povo, Trento, Italy

    \vspace*{10pt}

    $^\|$ emiliano.tolotti@unitn.it
\end{center}

\vspace*{15pt}

\begin{abstract}
\noindent
Ensemble methods in machine learning aim to improve prediction accuracy by combining multiple models. 
This is achieved by ensuring diversity among predictors to capture different data aspects. 
Homogeneous ensembles use identical models, achieving diversity through different data subsets, 
and weighted-average ensembles assign higher influence to more accurate models through a weight learning procedure. 
We propose a method to achieve a weighted homogeneous quantum ensemble 
using quantum classifiers with indexing registers for data encoding. 
This approach leverages instance-based quantum classifiers, 
enabling feature and training point subsampling through superposition and controlled unitaries,
and allowing for a quantum-parallel execution of diverse internal classifiers with different data compositions in superposition. 
The method integrates a learning process involving circuit execution and classical weight optimization, 
for a trained ensemble execution with weights encoded in the circuit at test-time. 
Empirical evaluation demonstrate the effectiveness of the proposed method, offering insights into its performance.

\vspace*{10pt}
\end{abstract}

\setcounter{footnote}{0}
\renewcommand{\thefootnote}{\alph{footnote}}

\vspace*{1pt}\textlineskip    

\section{Introduction}
Ensemble methods are well-known in machine learning and they aim to enhance prediction accuracy 
by combining the outputs of multiple models \cite{polikar_ensemble_2006}. 
One important principle of ensemble methods is to ensure diversity among the predictors, 
allowing each of them to capture different aspects of the data. 
As a result, the combined models provide a more comprehensive understanding and better prediction capability. 
In the context of homogeneous ensembles, where the internal models are of the same kind, 
diversity can be introduced by using different sets of data for each classifier. 
The main methods include bagging \cite{breiman_bagging_1996}, 
which considers subsets of training points sampled with replacement, 
and attribute bagging (random subspaces) that uses random subsets of features \cite{ho_random_1998}. 
Weighted-average ensembles are a subclass of ensemble methods 
where certain models are considered to be more accurate than others. 
These superior models receive higher weights through a weight learning procedure, 
giving them greater influence on the final prediction. 
They can be considered as stacking ensembles \cite{wolpert_stacked_1992}, 
where the stacked classifier performs the weighted aggregation. 
By combining the predictions of individual models with varying weights, 
weighted-average ensembles leverage the strengths of each model to achieve superior performance. 

The advantages of ensemble methods in the quantum setting have been presented in multiple works, 
where the proposed quantum-classical hybrid schemes show improvements in prediction performance 
\cite{zhang_efficient_2023, qin_improving_2022, incudini_resource_2023, west_boosted_2023, tolotti_ensembles_2024}. 
The idea of quantum ensembles has been introduced in \cite{schuld_quantum_2018, abbas_quantum_2020}, 
where the authors propose a Bayesian Model Averaging method with non-trained classifiers weighted by accuracy. 
The authors in \cite{macaluso_quantum_2022, macaluso_efficient_2024} propose an ensemble strategy 
with the introduction of diversity by action of a controlled unitary execution acting on the data register. 
However, their approach involves a single example with qubit encoding, 
where individual data instances are encoded in different qubits, 
and the introduction of diversity by unitary application is performed 
entangling different training observations for each control state. 

In contrast, we propose a method to achieve a weighted homogeneous quantum ensemble 
that employs a more efficient data encoding strategy based on instance-based quantum classifiers 
that rely on an indexing register, such as the quantum distance classifier \cite{schuld_implementing_2017}, 
the quantum cosine classifier \cite{pastorello_quantum_2021}, 
and a classification routine based on the SWAP test \cite{buhrman_quantum_2001}. 
These classifiers consider training instance vectors in superposition, 
requiring a different mechanism for introducing diversity. 
Additionally, rather than relying solely on diverse classifiers, which require independence to be advantageous, 
we propose a weighted aggregation of the internal models to increase accuracy. 

Building on this, our method introduces diversity by allowing subsampling of both features 
and training points for these instance-based quantum classifiers. 
This is achieved through permutation of indexes and post-measurement state selection. 
By exploiting superposition and an additional control register that determines the size of the ensemble, 
the method enables quantum-parallel execution of internal classifiers. 
To facilitate effective weighting, our approach requires two distinct execution modalities, 
one for training and one for testing, differing in the measured qubits. 
We evaluated the performance of the proposed method using benchmarks on real-world datasets, 
which demonstrated improved accuracy over single classifiers, 
showcasing the effectiveness of the parallel weighted aggregation of diverse classifiers.

The remainder of this paper is organized as follows: Section \ref{sec:class} provides 
background information on the quantum classifiers used in our approach. 
Section \ref{sec:prop} details our proposed method, 
including the quantum circuit design and the weight learning procedure. 
Section \ref{sec:exp} presents experimental results demonstrating the effectiveness of our method. 
Section \ref{sec:disc} discusses the implications of our findings and the role of the proposed ensemble. 
Finally, Section \ref{sec:concl} draws some conclusions and suggests directions for future research.

\section{Quantum classifiers}\label{sec:class}
In this section we provide a brief overview of the quantum instance-based binary classifiers considered in this work. 
All the classifiers consider amplitude encoding of the data vectors, 
specifically they rely on a data register encoding the training set defined as follows:
\begin{equation}
    \ket{\psi_d} = \frac{1}{\sqrt{N}}\sum_{i=0}^{N-1}\ket{i}\sum_{j=0}^{M-1}\xtrain_i^j\ket{j},
    \label{eq:enc}
\end{equation}
with
\begin{equation}
    \sum_{j=0}^{M-1}||\xtrain_i^j||^2=1, \quad \forall i \in \{0,...,N-1\},
    \label{eq:enc1}
\end{equation}
where $N$ and $M$ indicate the number of data instances and features respectively, 
$\xtrain_i^j$ indicates the feature $j$ of the $i$-th data instance and $\ket{i}$, $\ket{j}$ are the basis states 
corresponding to the binary representation of $i$, $j$. Feature vectors are unit-norm normalized. 
All the classifiers consider a basis encoding for the binary labels $\ytrain_i \in \{-1, +1\}$, mapped to the domain $\{1, 0\}$ as
\begin{equation}
    l_i = \frac{1 - \ytrain_i}{2}.
    \label{eq:label-to-qubit-state}
\end{equation}
\subsection{Quantum cosine classifier}\label{sec:class-cos}
The quantum cosine classifier \cite{pastorello_quantum_2021} is a binary classification algorithm 
that is based on the cosine similarity between feature vectors. Its prediction function is defined as
\begin{equation}
    \ytest(\xtest) = sign \left(\sum_{i=0}^{N-1}{\ytrain_i \cos{(\xtrain_i, \xtest)}}\right),
    \label{eq:cos-class-classical-formula}
\end{equation}
with $N$ being the number of training instances $\xtrain_i$ with respective binary labels $\ytrain_i$.
Its initial state is defined as
\begin{equation}
    \ket{\Psi} = \frac{1}{\sqrt{2}} (\ket{\psi_x}\ket{0} + \ket{\psi}\ket{1}) 
    \in \mathcal{H}_n \otimes \mathcal{H}_m \otimes \mathcal{H}_l \otimes \mathcal{H}_a ,
\end{equation}
with 
\begin{align*}
    \ket{\psi_x} &= \frac{1}{\sqrt{N}} \sum_{i=0}^{N-1}{\ket{i}\ket{\xtrain_i}\ket{l_i}} \in \mathcal{H}_n \otimes \mathcal{H}_m \otimes \mathcal{H}_l,\\
    \ket{\psi} &= \frac{1}{\sqrt{N}} \sum_{i=0}^{N-1}{\ket{i} \ket{\xtest} \ket{-}} \in \mathcal{H}_n \otimes \mathcal{H}_m \otimes \mathcal{H}_l .
\end{align*}
The classifier requires the execution of a SWAP test \cite{buhrman_quantum_2001} between an auxiliary qubit initialized in $\ket{+}$ 
and the ancilla $a$ of the initial state $\ket{\Psi}_a$. 

\begin{figure}[h!]
    \centering
    \begin{quantikz}
    \lstick{$\ket{0}$} & \gate{H} & \ctrl{2} &  \gate{H} & \meter{} & \setwiretype{c} \\
    \lstick{$\ket{+}$} & & \swap{1} & & & \\
    \lstick[2]{$\ket{\Psi}$} & & \targX{} & & & \rstick{$a$}\\
    & \qwbundle{} & & & & \rstick{$n,m,l$}
    \end{quantikz}
\end{figure}

The probability of measuring $1$ on the ancillary qubit controlling the SWAP test is
\begin{equation}
    \mathbb{P}(1) = \frac{1}{4}\left(1 - \frac{2\sum_{i=0}^{N-1}\ytrain_i\braket{\xtest|\xtrain_i}}{\sqrt{2}\sum_{i=0}^{N-1}||\xtest||^2+||\xtrain_i||^2}\right).
    \label{eq:cos_p}
\end{equation}
With unit-norm data vectors, since
\begin{equation}
\cos{(x, y)} = \frac{x \cdot y}{\lVert x\rVert \lVert y\rVert},
\end{equation}
the prediction according to Eq. \ref{eq:cos-class-classical-formula} is given by
\begin{equation}
    \ytest(\xtest) = sign(1-4\mathbb{P}(1)).
\end{equation}
\subsection{Quantum distance classifier}\label{sec:class-dist}
The quantum distance classifier proposed in \cite{schuld_implementing_2017} is a binary classification algorithm 
based on the squared Euclidean distance between feature vectors. Its prediction function is defined as
\begin{equation}
    \ytest(\xtest) = sign \left(\sum_{i=0}^{N-1}{\ytrain_i (1 - \frac{1}{4}||\xtest - \xtrain_i||^2})\right),
    \label{eq:dist-class-classical-formula}
\end{equation}
where $N$ is the number of training instances $\xtrain_i$ with binary labels $\ytrain_i$.
The initial state is defined as
\begin{equation}
    \ket{\Psi} = \frac{1}{\sqrt{2N}} \sum_{i=0}^{N-1} \ket{i}(\ket{0}\ket{\xtest} + \ket{1}\ket{\xtrain_i})\ket{l_i} 
    \in \mathcal{H}_n \otimes \mathcal{H}_a \otimes \mathcal{H}_m \otimes \mathcal{H}_l.
\end{equation}
The circuit is composed by a Hadamard gate acting on the ancilla $a$ and a conditional measurement 
of the label qubit $l$ after measuring the ancilla. 

\begin{figure}[h!]
    \centering
    \begin{quantikz}
    \lstick[3]{$\ket{\Psi}$} & \gate{H} & \meter{} & \setwiretype{c} & \rstick{$a$}\\
    & \qwbundle{} & & & \rstick{$n,m$}\\
    & & & \meter{} & \setwiretype{c} \rstick{$l$}
    \end{quantikz}
\end{figure}

Specifically, after a successful measurement of the ancilla qubit $a$ in 0, 
the probability of measuring $k\in\{0,1\}$ in $l$ can be written as
\begin{equation}
    \mathbb{P}(k|0) = \frac{1}{4Np_0} \sum_{i:l_i=k} ||\xtest + \xtrain_i||^2 = \frac{\sum_{i:l_i=k} ||\xtest + \xtrain_i||^2}{\sum_{i} ||\xtest + \xtrain_i||^2},
    \label{eq:dist_p}
\end{equation}
where $p_0$ is the probability of measuring $a$ in $0$.
With unit-norm data vectors, the following relationship holds:
\begin{equation}
    \frac{1}{4N} \sum_{i} ||\xtest + \xtrain_i||^2 = \frac{1}{N} \sum_{i} (1 - \frac{1}{4}||\xtest - \xtrain_i||^2),
\end{equation}
then, the predicted label according to Eq. \ref{eq:dist-class-classical-formula} is given by
\begin{equation}
    \ytest(\xtest) = sign\left(\mathbb{P}(0|0) - \frac{1}{2}\right).
\end{equation}

\subsection{Quantum SWAP test classifier}\label{sec:swap-test-classifier}
The authors in \cite{blank_quantum_2020} introduced a binary classification algorithm 
utilizing the SWAP test \cite{buhrman_quantum_2001}. 
This approach leverages a quantum kernel based on state fidelity. 
In our analysis, we consider a circuit that uses a single copy of the test data. 
However, this method can be extended to incorporate any number of copies, $n$, 
resulting in a kernel based on the $n$-th power of the fidelity. 
The prediction function is defined as
\begin{equation}
    \ytest(\xtest) = sign \left(\sum_{i=0}^{N-1}{\ytrain_i \cos^2{(\xtrain_i, \xtest)}}\right).
    \label{eq:swap-class-classical-formula}
\end{equation}
The initial state is defined as
\begin{equation}
    \ket{\Psi}\ket{\xtest} = \frac{1}{\sqrt{N}}\sum_{i=0}^{N-1} \ket{i}\ket{\xtrain_i}\ket{l_i}\ket{\xtest} 
    \in \mathcal{H}_n \otimes \mathcal{H}_m \otimes \mathcal{H}_l \otimes \mathcal{H}_x,
\end{equation}
where $\ket{x}$ indicates the state encoding the target vector $x$. 
The circuit applies the SWAP test between the qubits of the $m$ and $x$ registers, 
followed by a measurement of the control qubit of the SWAP $a$, and the label qubit $l$.

\begin{figure}[h!]
    \centering
    \begin{quantikz}
    \lstick{$\ket{0}$} & \gate{H} & \ctrl{4} &  \gate{H} & \meter{} & \setwiretype{c} \rstick{$a$}\\
    \lstick[3]{$\ket{\Psi}$} & \qwbundle{} & & & & \rstick{$n$}\\
    & \qwbundle{} & \targX{} & & & \rstick{$m$}\\
    & & & & \meter{} & \setwiretype{c} \rstick{$l$}\\
    \lstick{$\ket{x}$} & \qwbundle{} & \targX{} & & & \rstick{$x$}
    \end{quantikz}
\end{figure}

The joint probability of measuring $q$ in the ancillary qubit and $k$ in the label qubit can be written as
\begin{equation}
    \mathbb{P}(q,k) = \frac{1}{2N} \sum_{i:l_i=k} ||\xtest||^2||\xtrain_i||^2+(-1)^q|\braket{\xtest|\xtrain_i}|^2.
    \label{eq:swap_p}
\end{equation}
The classifier's output is computed from the expectation value of the observable 
corresponding to the measurement in the computational basis of the output qubits, as
\begin{equation}
    \braket{\sigma_z^a \otimes \sigma_z^l} = \mathbb{P}(0,0) - \mathbb{P}(0,1) - \mathbb{P}(1,0) + \mathbb{P}(1,1) = 
    \frac{1}{N} \sum_{i=0}^{N-1} \ytrain_i |\braket{\xtest|\xtrain_i}|^2.
\end{equation}
Considering unit-norm data vectors, the prediction according to Eq. \ref{eq:swap-class-classical-formula} is given by
\begin{equation}
    \ytest(\xtest) = sign(\mathbb{P}(0,0) - \mathbb{P}(0,1) - \mathbb{P}(1,0) + \mathbb{P}(1,1)).
\end{equation}

\subsection{General formulation of the initial state}
The classifiers described above share a common structure for the initial state, 
which we can express into a unified formulation. 
To provide a common notation for the different initializations,
we express the initial states in a more general form that highlights the data encoding. 
We define
\begin{equation}\label{eq:0}
    \ket{\psi}_{class} = \sum_{i,j=0}^{2^n-1,2^m-1} \alpha_{i,j}\ket{i}\ket{j}\ket{\psi_i} 
    = \sum_{i,j=0}^{N-1,M-1} \alpha_{i,j}\ket{i}\ket{j}\ket{\psi_i} 
    \in \mathcal{H}_n \otimes \mathcal{H}_m \otimes \mathcal{H},
\end{equation}
where we consider $n$ and $m$ qubits to encode the training samples and features respectively, so that $N\leq2^n$ and $M\leq2^m$.
The coefficient $\alpha_{i,j}$ represents the amplitude of the $j$-th feature of the $i$-th data point, 
so $\alpha_{i,j}=0, \, \forall i,j \notin [0, N-1]\times[0, M-1]$.
In this expression, each classifier encodes a dataset of $N$ training instances in superposition, 
with each instance represented by an indexing register ($\mathcal{H}_n$), identifying the training sample, 
and a feature register ($\mathcal{H}_m$) that encodes feature vectors through amplitude encoding. 
The auxiliary register ($\mathcal{H}$) stores class labels, ancillas, or other classifier-specific information.

We can express the initial state of the quantum classifiers in the unified form,
by determining specific values for the amplitudes $\alpha_{i,j}$ and the auxiliary register state $\ket{\psi_i}$.
For instance, the initial states of the quantum cosine and distance classifiers (referred to as \textit{cdc}), 
described in Section \ref{sec:class-cos} and Section \ref{sec:class-dist} respectively, have a form of this type:
\begin{equation}\label{eq:-1a}
    \ket{\Psi^{\text{cdc}}} = \frac{1}{\sqrt{2N}}\sum_{i=0}^{N-1} \ket{i}(\ket{\xtrain_i}\ket{\gamma}\ket{\phi_i} + \ket{\xtest}\ket{1-\gamma}\ket{\theta_{i}}) 
    \in \mathcal{H}_n \otimes \mathcal{H}_m \otimes \mathcal{H}_a \otimes \mathcal{H}_c,
\end{equation}
where the input sample $x$ is encoded in the same register of the training samples 
but entangled with different values $\gamma \in \{0,1\}$ of the ancilla qubit $a$,
and $\ket{\phi_i}, \ket{\theta_{i}}$ encode classifier-specific information.
To match the general form, the initial states of these classifiers can be rewritten as
\begin{equation}\label{eq:0a}
    \ket{\Psi^{\text{cdc}}} = \frac{1}{\sqrt{2N}}\sum_{i,j=0}^{N-1,M-1} \ket{i}\ket{j}(\xtrain_i^j \ket{\gamma}\ket{\phi_i} + \xtest^j \ket{1-\gamma}\ket{\theta_{i}})
    = \ket{\psi}_{class},
\end{equation}
where the amplitudes and the auxiliary register state are identified as
\begin{equation}
    \alpha_{i,j}\ket{\psi_i} = \frac{1}{\sqrt{2N}} (\xtrain_i^j \ket{\gamma}\ket{\phi_i} + \xtest^j \ket{1-\gamma}\ket{\theta_{i}}), 
    \quad \forall i \in \{0,...,N-1\}, \forall j \in \{0,...,M-1\}.
\end{equation}

The SWAP test classifier, described in Section \ref{sec:swap-test-classifier}, 
encodes the test point $x$ in an additional register ($\mathcal{H}_x$) instead, 
and the initial state can be generally written as follows:
\begin{equation}\label{eq:-1b}
    \ket{\Psi^{\text{swap}}} = \frac{1}{\sqrt{N}}\sum_{i=0}^{N-1} \ket{i}\ket{\xtrain_i}\ket{\psi_i}\ket{\xtest}
    \in \mathcal{H}_n \otimes \mathcal{H}_m \otimes \mathcal{H}_c \otimes \mathcal{H}_x.
\end{equation}
This state can be recast into the unified form as
\begin{equation}\label{eq:0b}
    \ket{\Psi^{\text{swap}}} = \frac{1}{\sqrt{N}}\sum_{i,j=0}^{N-1,M-1} \xtrain_i^j \ket{i}\ket{j}\ket{\psi_{i}} \left(\sum_{z=0}^{M-1} \xtest_z \ket{z}\right)
    = \ket{\psi}_{class} \otimes \ket{\xtest},
\end{equation}
with
\begin{equation}
    \alpha_{i,j}\ket{\psi_i} = \frac{1}{\sqrt{N}} \xtrain_i^j \ket{\psi_i},
    \quad \forall i \in \{0,...,N-1\}, \forall j \in \{0,...,M-1\}.
\end{equation}

\section{Proposed ensemble method}\label{sec:prop}
In this section we outline the procedure and the details of the proposed quantum ensemble scheme. 

\subsection{Procedure}
The procedure considers classifiers of the type reported in Section \ref{sec:class},
whose initial states are expressed using the unified notation in Eqs. \ref{eq:0a}, \ref{eq:0b}.
For presenting the procedure, we utilize the general form in Eq. \ref{eq:0},
to which we add a control register and an additional ancilla qubit.
The size ($d$) of the control register determines the size of the ensemble, i.e. the number of diverse internal classifiers ($2^d$).
The general initial state for the proposed ensemble, considering the control register, the initialized classifier and the additional ancilla qubit can be written as follows:
\begin{equation}\label{eq:1}
    \ket{\Phi_0} = \ket{0}_{ctrl}^{\otimes d} \otimes \ket{\psi}_{class} \otimes \ket{0}_{aux} = 
    \ket{0}^{\otimes d} \sum_{i,j=0}^{2^n-1,2^m-1} \alpha_{i,j}\ket{i}\ket{j}\ket{\psi_i} \ket{0}.
\end{equation}
This general representation simplifies the description of the ensemble's operations,
and will be reconducted back to the classifier-specific formulations when it will be needed.

We discuss the steps of the procedure at test-time in Section \ref{sec:test-time}, 
followed by the modifications utilized for training in Section \ref{sec:train-time}.

\subsubsection{Quantum ensemble circuit at test time}\label{sec:test-time}
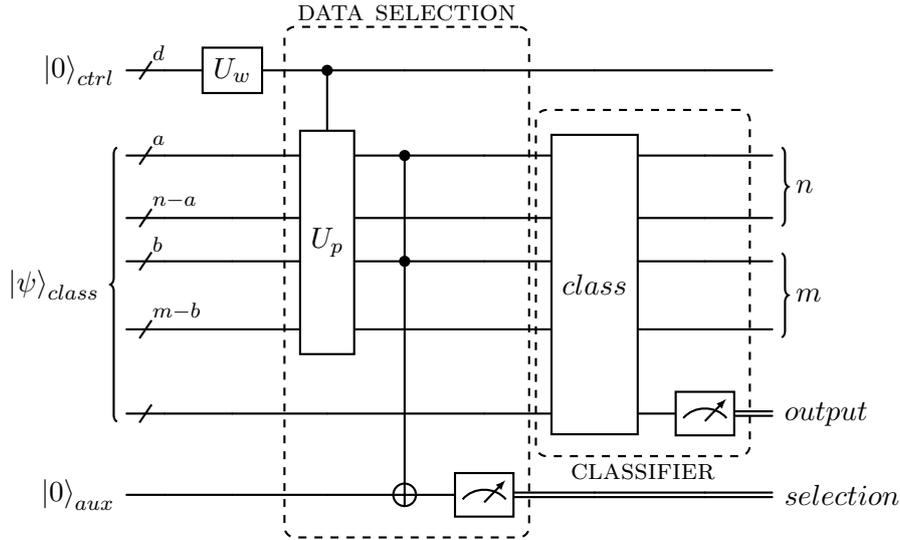
\begin{figure}[t!]
    \centering
    \begin{quantikz}
    \lstick{$\ket{0}_{ctrl}$} & \qwbundle{d} & \gate{U_w} & \ctrl{2} 
    \gategroup[7,steps=3,style={dashed,rounded corners, inner xsep=2pt},background,label style={label position=above,anchor=north,yshift=0.2cm}]
    {{\sc data selection}} & & & & & \\
    \lstick[5]{$\ket{\psi}_{class}$} & \qwbundle{a} & & \gate[4]{U_p} & \ctrl{2} & & \gate[5]{class} 
    \gategroup[5,steps=2,style={dashed,rounded corners, inner xsep=2pt},background,label style={label position=below,anchor=north,yshift=-0.15cm}]
    {{\sc classifier}} & & \rstick[2]{$n$}\\
    & \qwbundle{n-a} & & & & & & & \\
    & \qwbundle{b} & & & \ctrl{3} & & & & \rstick[2]{$m$}\\
    & \qwbundle{m-b} & & & & & & & \\
    & \qwbundle{} & & & & & & \meter{} & \setwiretype{c} \rstick{$output$}\\
    \lstick{$\ket{0}_{aux}$} & & & & \targ{} & \meter{} & \setwiretype{c} & & \rstick{$selection$}
    \end{quantikz}
    \caption{Circuit diagram of the ensemble at test time. 
    The control register is prepared with the weights encoded in the amplitudes. 
    Then, the controlled permutation unitary is executed, acting on the data register of the classifier. 
    The data register then partially controls ($a$ qubits out of $n$ and $b$ qubits out of $m$) a NOT gate acting on the ancilla qubit, 
    which is subsequently measured for data selection. 
    The classifier is executed, and the related output qubit is measured.
    Specifically, the displayed circuit is designed to execute a quantum classifier of type \textit{cdc}.}
    \label{fig:circuit-test}
\end{figure}

The test-time procedure performs the ensemble's prediction for an input $\xtest$, 
by executing the circuit exemplified in the diagram in Figure \ref{fig:circuit-test}.
Suppose we have access to a positive unit-norm normalized weight vector $w$ for aggregation of the internal classifiers of the ensemble. 
We encode the weights related to each internal classifier into the amplitudes of the control register 
by action of the unitary $U_{w}$:
\begin{equation}\label{eq:2}
    \ket{\Phi_1} = U_{w}\ket{\Phi_0} = \sum_{c=0}^{2^d-1}\sqrt{w_c}\ket{c} \sum_{i,j=0}^{2^n-1,2^m-1} \alpha_{i,j}\ket{i}\ket{j}\ket{\psi_i} \ket{0},
\end{equation}
where $\sqrt{w_c}$ indicates the amplitude related to the weight $w_c$. 

In the next step, we apply controlled ``permutation" unitary logic $U_p$, 
controlled by the control register and targeting the indexing register of training points and features. 
The purpose of the $U_p$ unitary is to permute the elements of the training set in superposition,
by entangling the data register with the control register. 
This can be achieved through bit-flip logic, as an application of CNOT and CSWAP gates. 
After the bit flipping, a different permutation of training samples and features is entangled 
with each basis state of the control register.
The action of $U_p$ can be written as follows:
\begin{equation}\label{eq:3}
    \ket{\Phi_2} = U_{p}\ket{\Phi_1} = \sum_{c=0}^{2^d-1}\sqrt{w_c}\ket{c} \sum_{i,j=0}^{2^n-1,2^m-1} \alpha^c_{i,j}\ket{i}\ket{j}\ket{\psi_i} \ket{0},
\end{equation}
where the amplitude $\alpha^c_{i,j}$ indicates a different feature amplitude depending on the control $c$ for each index $i,j$. 

After the permutation, a (multi-)controlled-NOT (MCX) gate is applied with control on partial index and feature registers, 
while acting on the ancillary qubit, 
\begin{equation}\label{eq:4}
    \ket{\Phi_3} = U_\text{MCX}\ket{\Phi_2} = \sum_{c=0}^{2^d-1}\sqrt{w_c}\ket{c} 
    \left(\sum_{i,j\in sel} \alpha^c_{i,j}\ket{i}\ket{j}\ket{\psi_i} \ket{0} + 
    \sum_{i,j\notin sel} \alpha^c_{i,j}\ket{i}\ket{j}\ket{\psi_i}\ket{1}\right).
\end{equation}
Referring to Figure \ref{fig:circuit-test}, assuming for example control by 2 qubits ($a=b=1$), 
specifically the least significant in binary notation of the index and feature registers respectively, 
the action of the Toffoli gate (CCX) is
\begin{equation}\label{eq:4a}
    \begin{aligned}
    \ket{\Phi_3'} = U_\text{CCX}\ket{\Phi_2} = \sum_{c=0}^{2^d-1}\sqrt{w_c}\ket{c} \sum_{i,j=0}^{2^{n-1}-1,2^{m-1}-1} 
    &\alpha^c_{2i,2j}\ket{2i}\ket{2j}\ket{\psi_{2i}}\ket{0} \\
    + &\alpha^c_{2i+1,2j}\ket{2i+1}\ket{2j}\ket{\psi_{2i+1}}\ket{0}\\
    + &\alpha^c_{2i,2j+1}\ket{2i}\ket{2j+1}\ket{\psi_{2i}}\ket{0} \\
    + &\alpha^c_{2i+1,2j+1}\ket{2i+1}\ket{2j+1}\ket{\psi_{2i+1}}\ket{1}.
    \end{aligned}
\end{equation}

The selection of training indexes and features is obtained by measuring the ancilla in the state $\ket{\Phi_3}$, 
obtaining a post-measurement state that reflects the corresponding state of the control qubits. 
Measuring 1 selects the training indexes and features whose binary representations 
have 1 in correspondence with the qubits that controlled the action on the ancilla, while measuring 0 selects the complementary set. 
The measurement probability is expected to be proportional to the size of the selected subset, 
assuming that the feature amplitudes are uniformly distributed.
Measuring the ancilla collapses the state to a superposition of multiple diverse datasets, 
where each basis state of the control register is entangled with a different encoded data composition.
The post-measurement state results to be
\begin{equation}\label{eq:5}
    \ket{\Phi_4} = \frac{1}{\sqrt{p_0}}\sum_{c=0}^{2^d-1}\sqrt{w_c}\ket{c} \sum_{i,j\in sel} \alpha^c_{i,j}\ket{i}\ket{j}\ket{\psi_i} \ket{0},
\end{equation}
where $p_0$ is the probability of measuring 0 in the ancilla qubit.
Supposing we want to select the set where the ancilla is 0 
in the $\ket{\Phi_3'}$ state ($p_0 \approx 0.75$), in the example above ($a=b=1$)
the resulting state after the data selection by measurement is as follows:
\begin{equation}\label{eq:5a}
    \begin{aligned}
    \ket{\Phi_4'} = \frac{1}{\sqrt{p_0}}\sum_{c=0}^{2^d-1}\sqrt{w_c}\ket{c} \sum_{i,j=0}^{2^{n-1}-1,2^{m-1}-1} 
    &\alpha^c_{2i,2j}\ket{2i}\ket{2j}\ket{\psi_{2i}}\ket{0} \\
    + &\alpha^c_{2i+1,2j}\ket{2i+1}\ket{2j}\ket{\psi_{2i+1}}\ket{0}\\
    + &\alpha^c_{2i,2j+1}\ket{2i}\ket{2j+1}\ket{\psi_{2i}}\ket{0}.
    \end{aligned}
\end{equation}
Note that in the case of cosine and distance classifiers (\textit{cdc}) $\ket{\Phi^{\text{cdc}}_4} = \ket{\Phi_4}$,
while in the case of a SWAP test (\textit{swap}) classifier with separated test point register we have
\begin{equation}
    \ket{\Phi^{\text{swap}}_4} = \frac{1}{\sqrt{p_0}}\sum_{c=0}^{2^d-1}\sqrt{w_c}\ket{c} \sum_{i,j\in sel} \alpha^c_{i,j}\ket{i}\ket{j}\ket{\xtest^{(c)}}\ket{\psi_i} \ket{0},
\end{equation}
where the selection is performed on the test point register as well. 
Here, the permutation unitary $U_p$ is applied also on $\ket{\xtest}$, performing the same feature permutation 
of the training points also on the test point, for each basis state of the control register.
In this case the circuit in Figure \ref{fig:circuit-test},
which is designed for the \textit{cdc} classifier, should include also the test point register,
and $U_p$ should act on both the training and test point registers.

A reformulation of the preceding results can shed light on the nature of the elaboration.
The equation Eq. \ref{eq:5} can be rewritten substituting back the original indexes and features, 
interpreting the selection as a different composition of the training set.
In the case of the quantum cosine and distance classifiers (Eq. \ref{eq:-1a}), the state after the data selection can be written as
\begin{equation}\label{eq:5-a}
    \ket{\Phi^{\text{cdc}}_4} = \frac{1}{\sqrt{p_0}}\sum_{c=0}^{2^d-1}\sqrt{w_c}\ket{c} \sum_{i \in sel} \delta_{c,i} \ket{i}(\ket{\xtrain_i^{(c)}}\ket{\gamma}\ket{\psi_i} + \ket{\xtest^{(c)}}\ket{1-\gamma}\ket{\phi_{i}})\ket{0},
\end{equation}
while in the case of the SWAP test classifier (Eq. \ref{eq:-1b}) it can be written as
\begin{equation}\label{eq:5-b}
    \ket{\Phi^{\text{swap}}_4} = \frac{1}{\sqrt{p_0}}\sum_{c=0}^{2^d-1}\sqrt{w_c}\ket{c} \sum_{i \in sel} \delta_{c,i} \ket{i}\ket{\xtrain_i^{(c)}}\ket{\psi_i}\ket{\xtest^{(c)}}\ket{0}.
\end{equation}
In both cases, $\xtrain_i^{(c)}$ and $\xtest^{(c)}$ are the selected training and test input points respectively, 
after applying the same permutation and selection of features, 
where $\delta_{c,i}$ represent normalization coefficients for each control basis state and training sample.
It should be noted that the formulation resembles a superposition of the initial states of the classifiers,
with the control register entangled with different compositions of the training and the test points.


The original circuit (denoted \textit{class} in Figure~\ref{fig:circuit-test}) of the classifier (\textit{cdc} or \textit{swap}) is then executed:
\begin{equation}\label{eq:6}
    \ket{\Phi_5} = \sum_{c=0}^{2^d-1}\sqrt{w_c}\ket{c} \ket{\phi_c} \ket{\varphi_c^\text{out}}\ket{0},
\end{equation}
where $\ket{\varphi_c^\text{out}}$ denotes the output qubits state of each classifier, generally not normalized,
and $\ket{\phi_{c}}$ indicates the rest of each classifier's state. 
Considering an observable \( O \),
the output of the ensemble classifier is given by the expectation value of \( O \), 
obtained by collecting statistics of the measurement outcomes:
\begin{equation}\label{eq:7}
    \mathbb{E}[O] = \frac{\sum_{c=0}^{2^d-1} w_c \bra{\varphi_c^\text{out}} O \ket{\varphi_c^\text{out}}}{\sum_{c=0}^{2^d-1} w_c \braket{\varphi_c^\text{out} | \varphi_c^\text{out}}}
    = \sum_{c=0}^{2^d-1} w_c \bra{\varphi_c^\text{out}} O \ket{\varphi_c^\text{out}},
\end{equation}
which resembles a linear combination of the outputs $\bra{\varphi_c^\text{out}} O \ket{\varphi_c^\text{out}}$ of the internal classifiers,
where the weights are represented by the unit-norm vector $w$.

As a special case, considering a single output qubit and projective measurement on the computational basis, 
$\ket{\varphi_c^\text{out}} = \beta_0^c \ket{0} + \beta_1^c \ket{1}$ and $O = \ket{0}\bra{0}$, 
Eq. \ref{eq:7} reduces to the following linear combination of outcome probabilities:
\begin{equation}\label{eq:7-special}
    \mathbb{P}(0) = \frac{\sum_{c=0}^{2^d - 1} w_c |\beta_0^c|^2}{\sum_{c=0}^{2^d - 1} w_c \left(|\beta_0^c|^2 + |\beta_1^c|^2\right)}
    = \sum_{c=0}^{2^d-1}{w_c |\beta_0^c|^2}.
\end{equation}

\subsubsection{Quantum ensemble circuit at train time}\label{sec:train-time}

\begin{figure}[t!]
    \centering
    \begin{quantikz}
    \lstick{$\ket{0}_{ctrl}$} & \qwbundle{d} & \gate{H} & \ctrl{2} 
    \gategroup[7,steps=3,style={dashed,rounded corners, inner xsep=2pt},background,label style={label position=above,anchor=north,yshift=0.2cm}]
    {{\sc data selection}} & & & & \meter{} & \setwiretype{c} \rstick{$internal$}\\
    \lstick[5]{$\ket{\psi}_{class}$} & \qwbundle{a} & & \gate[4]{U_p} & \ctrl{2} & & \gate[5]{class} 
    \gategroup[5,steps=2,style={dashed,rounded corners, inner xsep=2pt},background,label style={label position=below,anchor=north,yshift=-0.15cm}]
    {{\sc classifier}} & & \rstick[2]{$n$}\\
    & \qwbundle{n-a} & & & & & & & \\
    & \qwbundle{b} & & & \ctrl{3} & & & & \rstick[2]{$m$}\\
    & \qwbundle{m-b} & & & & & & & \\
    & \qwbundle{} & & & & & & \meter{} & \setwiretype{c} \rstick{$output$}\\
    \lstick{$\ket{0}_{aux}$} & & & & \targ{} & \meter{} & \setwiretype{c} & & \rstick{$selection$}
    \end{quantikz}
    \caption{Circuit diagram of the ensemble at training time. 
    The control register is prepared in a balanced superposition using Hadamard gates. 
    Then, the controlled permutation unitary is executed, acting on the data register of the classifier. 
    The data register then partially controls a NOT gate acting on the ancilla qubit, 
    which is subsequently measured for data selection. 
    The classifier is executed, and finally, 
    the control register is measured to get an internal classifier with the related output.
    Specifically, the displayed circuit is designed to execute a quantum classifier of type \textit{cdc}.}
    \label{fig:circuit-train}
\end{figure}
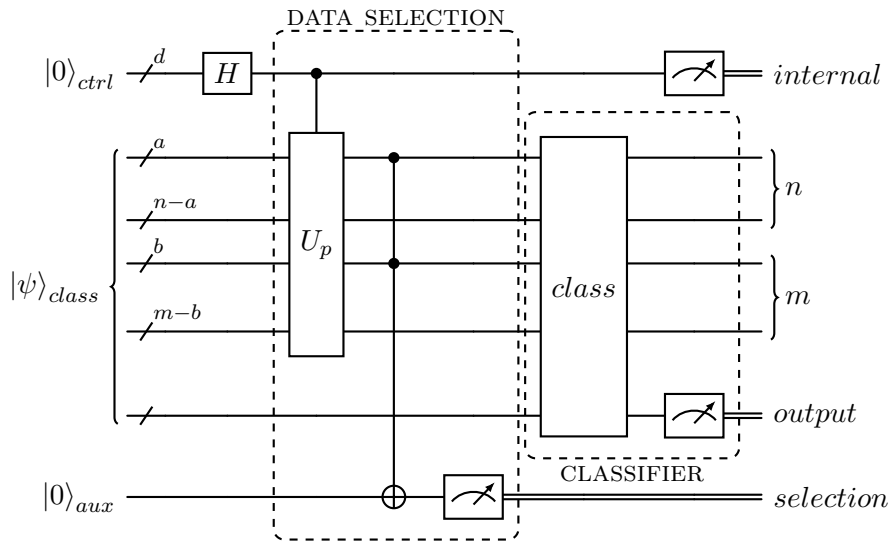

The training procedure provides a way to obtain the weight vector $w$ to be used at test time. 
To achieve this, it considers a set of samples as input and collects the outputs of each of the internal classifiers, 
which are used to determine the weights. The training circuit is similar to that during testing, 
with the corresponding diagram reported in Figure \ref{fig:circuit-train}, where the differences lie in the control register. 
During training, the classifiers are initialized on the training set and we consider uniform aggregation weights. 
The control register can be prepared in balanced superposition by applying Hadamard gates, 
replacing $U_w$ with $H^{\otimes d}$ in Eq. \ref{eq:2}. The same shuffling logic is then executed (Eqs. \ref{eq:3}, \ref{eq:4}), 
and a measurement is taken to select the same data in superposition (Eq. \ref{eq:5}),
so the ancilla output needs to match that at test time.


Upon a successful postselection (e.g., ancilla qubit in state $\ket{0}$), the classifier is executed:
\begin{equation}\label{eq:6a}
    \ket{\Phi_5}_{\text{tr}} = \frac{1}{2^{d/2}} \sum_{c=0}^{2^d-1} \ket{c} \ket{\phi_c} \ket{\varphi_c^\text{out}} \ket{0}.
\end{equation}
The training phase requires measuring the control register to collect 
the predictions of the individual classifiers on a validation dataset. 
After measurement, the state collapses to that of a randomly chosen internal classifier $c$, 
determined by the measurement output of the control register:
\begin{equation}\label{eq:6b}
    \ket{\Phi_6}_{\text{tr}} = \frac{1}{\sqrt{p_c}} \ket{c} \ket{\phi_c} \ket{\varphi_c^\text{out}} \ket{0},
\end{equation}
where $p_c$ is the probability of measuring the control register in the basis state corresponding to the classifier $c$,
which is proportional to the norm of its output state:
\begin{equation}\label{eq:6c}
    p_c = \frac{\braket{\varphi_c^\text{out} | \varphi_c^\text{out}}}{2^d}.
\end{equation}
Then, the observable $O$ is measured on the output qubits of $\ket{\varphi_c^\text{out}}$.
The prediction of the single internal classifier $c$ is then:
\begin{equation}\label{eq:7a}
    \mathbb{E}[O]_c = \frac{\bra{\varphi_c^\text{out}} O \ket{\varphi_c^\text{out}}}{\braket{\varphi_c^\text{out} | \varphi_c^\text{out}}}.
\end{equation}
These values are estimated by executing the circuit multiple times,
and are used to compute the aggregation weights $w_c$.

\begin{algorithm}[t!]
    \small
    \caption{Hybrid training procedure}\label{alg:training}
    \KwIn{validation set $D_{val}(x, y)$, training circuit $QC_{tr}$}
    \KwOut{aggregation weights $w$}
    \For{$x_i \in D_{val}(x, y)$}{
        \For{$c \gets 0$ \textbf{to} $2^d-1$}{
            $k_0[c] \gets 0$\tcp*{init 0 counts for output qubit of each classifier}
            $k_1[c] \gets 0$\tcp*{init 1 counts for output qubit of each classifier}
        }
        \For{$s \gets 0$ \textbf{to} $N_{shots}-1$}{
            $c^s_i, o^s_i \gets QC_{tr}(x_i)$\tcp*{run the circuit for sample $x_i$}
            \eIf(\tcp*[f]{accumulate counts for classifier $c^s_i$}){$o^s_i = 0$}
            {
                $k_0[c^s_i]++$\;
            }{
                $k_1[c^s_i]++$\;
            }
        }
        \For{$c \gets 0$ \textbf{to} $2^d-1$}{
            $p[i,c] \gets (k_0[c] + k_1[c]) / N_{shots}$\tcp*{estimate probability $p_c$ (Eq. \ref{eq:6c})}
            $p_0[i,c] \gets k_0[c] / (k_0[c] + k_1[c])$\tcp*{estimate probability $\mathbb{P}(0)_c$ (Eq. \ref{eq:7a})}
        }
    }
    $w \gets \textit{weights\_learning}(p, p_0)$\tcp*{compute weights $w$}
    \Return $w$\;
\end{algorithm}

The training circuit execution is reported in Algorithm \ref{alg:training}, 
which can be applied to predict the training set (or a different validation). 
For simplicity, here we assume the output of the internal classifiers to be a single qubit,
but the procedure can be generalized to multiple output qubits and observables.
In this setting, $O = \ket{0}\bra{0}$ is the observable for measuring the output qubit.
The expectation $\mathbb{E}[O]_c$ for the internal classifier $c$,
corresponds to the probability $\mathbb{P}(0)_c$ of measuring 0 in the output qubit.
The procedure estimates the values in Eqs. \ref{eq:6c}, \ref{eq:7a}, 
for each internal classifier $c$ and for each validation sample $x_i$,
that are used to compute the weights $w$ for the aggregation of the internal classifiers.
The algorithm iterates over the input samples $x_i$ and runs the training circuit 
for a determined number of shots $N_{shots}$, related to the quality of the estimation. 
Each circuit execution outputs a random classifier index $c^s_i$ and an output measurement $o^s_i$ for that classifier and current data sample $i$. 
Here we assume the circuit to output a successful realization, but in practice
the data selection circuit may fail, and the procedure should be repeated until a successful measurement is obtained.
The results are then accumulated, for each internal classifier and validation sample, and the output values are estimated. 
Afterwards, the individual outputs are used to calculate the weights $w$ for the classifiers to be used at test-time. 
These weights can be computed arbitrarily with classical or quantum procedures. 
A possibility is applying a weighting based on their performance on the training set, 
such as weighting by accuracy or by considering exponential weight decaying with errors. 
The chosen method involves optimization using logistic regression. 
The computed aggregation weights are then encoded in the amplitudes of the control register, 
taking into account positive weights. This enables us to consider an accuracy-weighted ensemble, 
having measured the performance in predicting the training set, or to tune the weight vector with a model fit.

\subsection{Data selection}

\newcommand{\arrayindex}[2]{\textit{#1}$[\mathit{#2}]$}
\newcommand{\arrayindexmath}[2]{\textit{#1}$[$#2$]$}
\newcommand{\ctr}[1]{\textbf{\smaller{control}}(#1)}
\newcommand{\tgt}[1]{\textbf{\smaller{target}}(#1)}
\newcommand{\mgate}[1]{\textbf{#1}}
\begin{algorithm}[t!]
    \small
    \caption{Data selection quantum circuit}\label{alg:selection}
    \KwIn{control qubits, index qubits, features qubits, ancilla}
    \KwOut{circuit producing $2^\mathit{control\_qubits}$ different data sets with $\approx \frac{3}{4}$ original size in superposition}
    \For{$\mathit{idx} \gets 0$ \textbf{to} $\mathit{control\_qubits} - 1$}{
        \If{$\mathit{idx} < \lfloor index\_qubits/2 \rfloor$}{
            Apply \mgate{CSWAP} \ctr{\arrayindex{control}{idx}} \tgt{\arrayindex{index}{idx}, \arrayindexmath{index}{$\mathit{idx}+\lfloor \mathit{index\_qubits}/2 \rfloor$}}\;
        }
        \If{$idx < \lfloor feature\_qubits/2 \rfloor$}{
            Apply \mgate{CSWAP} \ctr{\arrayindex{control}{idx}} \tgt{\arrayindex{feature}{idx}, \arrayindexmath{feature}{$\mathit{idx}+\lfloor \mathit{feature\_qubits}/2 \rfloor$}}\;
        }
        \If{$idx < index\_qubits$}{
            Apply \mgate{CNOT} \ctr{\arrayindex{control}{idx}} \tgt{\arrayindex{index}{idx}}\;
        }
        \If{$idx < feature\_qubits$}{
            Apply \mgate{CNOT} \ctr{\arrayindex{control}{idx}} \tgt{\arrayindex{feature}{idx}}\;
        }
    }
    \For{$\mathit{idx} \gets 0$ \textbf{to} $\mathit{control\_qubits} - 1$}{
        \For{$\mathit{idx1} \gets 0$ \textbf{to} $\mathit{control\_qubits} - 1$}{
            \If{$\mathit{idx} \neq \mathit{idx1}$}{
                Apply \mgate{CSWAP} \ctr{\arrayindex{control}{idx}} \tgt{\arrayindex{index}{idx}, \arrayindex{index}{idx1}}\;
                Apply \mgate{CSWAP} \ctr{\arrayindex{control}{idx}} \tgt{\arrayindex{feature}{idx}, \arrayindex{feature}{idx1}}\;
            }
        }
    }
    Apply \mgate{CCX} \ctr{\arrayindexmath{index}{0},\arrayindexmath{feature}{0}} \tgt{\textit{ancilla}}\;
    $a \gets $ \textbf{measure} \textit{ancilla}\;
    \eIf{$a == 0$}{
        \textbf{continue}\;
    }
    {
        \textbf{goto} initial state preparation\;
    }
\end{algorithm}

The data selection procedure introduces diversity in the ensemble,
by selecting the data subsets of each internal classifier.
It is composed by a permutation of the dataset in superposition and a controlled action on an ancilla 
controlled by a few qubits of the data register with successive measurement of the ancilla. 
Regarding the permutation, it is performed by applying controlled operations acting on the data register 
controlled by the control register. This action, referred as $U_p$ in Eq. \ref{eq:3}, 
produces bit flips of the basis states of the data register changing its indexing differently 
for each basis state of the control register. 
The logic can be applied to both the index and feature registers of the classifier, 
achieving  different ordering for the training instances and features producing $2^d$ potentially different 
amplitude permutations entangled with a control register of size $d$. 
For classifiers that consider the test point in a separate register in the initial state, 
as in Eq. \ref{eq:-1b}, such as the the SWAP test classifier described in \ref{sec:swap-test-classifier}, 
the same controlled permutation logic used for the features needs to be applied also to the test register. 
In this way the different permutations of the features are consistent between training and test points.

To produce diversity for each classifier in superposition, 
we make the data register collapse to a subset of the indexes or features by measurement. 
We achieve that by a multi-controlled-NOT acting on an auxiliary qubit and controlled by a subset of the data register qubits 
(in Figs. \ref{fig:circuit-test}, \ref{fig:circuit-train}, $a$ and $b$ qubits of the index and feature register respectively), 
as described in Eq. \ref{eq:4}. 
The measurement of the auxiliary qubit collapses the states of the qubits that controlled the operation, 
effectively producing a subset of indexes of the data register, which will contain different data in superposition, 
as in Eq. \ref{eq:5}. We considered a single ancillary qubit for this process, 
but an alternative approach is to consider multiple qubits to refine the data selection. 
For instance, one ancilla could be employed for selecting training indexes while the other focuses on the features, 
with each ancilla controlled by either the index or feature qubits, for defining more complex data selection patterns.

The permutation unitary and the ancilla control with measurement outcome are chosen, 
and deterministically decide the ensemble's internal classifiers data compositions and size with a given training set order. 
Their action can be arbitrary, with the requirement that it would produce different data sets which the ensemble can leverage from. 
For example the permutation can include multiple CSWAP and CNOT gates acting between different qubits, 
while the index selection can be performed by considering different control qubits in the data register, 
and the measurement can be accepted if in 1, considering the subset of the indexes that activated the control, 
or in 0, considering the outer set. The success probability of this measurement, 
assuming the data is uniformly distributed, is proportional to the size of the subset we aim to select. 
Specifically, it approximately corresponds to the fraction of the subset relative to the total training set.

In this work we considered the data selection circuit described by the pseudocode in Algorithm \ref{alg:selection}, 
which employs CSWAP and CNOT operations for the controlled permutation, 
and for the selection the action of a CCNOT gate controlled by the first qubits of the index and feature registers 
and post-selection with 0 measurement.
The circuit is designed to select a subset with a size of $\approx \frac{3}{4}$ of the original training set,
with a success probability of $\approx \frac{3}{4}$.
Specifically, the circuit discards the second half of the features the from the second half of the training points 
that are permuted differently in each internal classifier.

\subsection{Weighted output combination}
This section explains how the outputs of the internal classifiers are combined to produce the final prediction of the ensemble.
At test time, the final ensemble prediction is the linear combination of the contributions $\beta$ of each internal classifier, 
weighted according to the positive unit-norm weight vector $w$, encoded in the amplitudes of the control register, 
according to Eq. \ref{eq:7}, which we can rewrite in vector notation as $w^T\beta$.
To allow for both positive and negative weights, 
the vector $w$ can be rescaled to be entirely positive by adding a bias $b$. 
In this case, the circuit needs to be executed twice for each input, 
with the rescaled positive weights $w+b$ and with uniform weights $b$ (unit-norm normalized),
and the final output is obtained by the difference of the runs, rescaled accordingly:
\begin{equation}
    \frac{w^T\beta}{||w||} = \frac{1}{||w||} \left(||w+b|| \frac{(w+b)^T\beta}{||w+b||} - ||b||\frac{b^T\beta}{||b||}\right).
\end{equation}

After obtaining the output probability of the classifier, 
by collecting statistics of the measurement outcome of the output qubit for multiple circuit repetitions, 
the output can be post-processed classically, obtaining the weighted ensemble output. 
In our case, the final decision is determined according to parameters computed with logistic regression.
Specifically, as detailed in Sec. \ref{sec:train-time}, during training the circuit is executed on a validation set to obtain
the estimations of $p_{c,i}$ and $\mathbb{E}[O]_{c,i}$,
for each internal classifier $c$ and validation sample $i$ (Eqs. \ref{eq:6c} and \ref{eq:7a}),
to be used for the parameters optimization.
Then, we express the output of the ensemble classifier $\mathbb{E}[O]_i$ for each validation sample $i$,
by means of the internal classifiers' outputs and weights $w$, by rewriting Eq. \ref{eq:7} as
\begin{equation}
    \mathbb{E}[O]_i = \frac{\sum_{c=0}^{2^d-1}{w_c \cdot p_{c,i} \cdot \mathbb{E}[O]_{c,i}}}{\sum_{c=0}^{2^d-1}{w_c \cdot p_{c,i}}}.
\end{equation}
The optimization minimizes the log-loss function:
\begin{equation}\label{eq:7b}
    \mathcal{L}(w, b, k) = -\frac{1}{|D_{val}|}\sum_{i=1}^{|D_{val}|} \left( y_i \log(\sigma(z_i)) + (1-y_i) \log(1-\sigma(z_i))\right),
\end{equation}
where $\sigma(z) = \frac{1}{1 + e^{-z}}$ is the sigmoid function, $z_i = k \mathbb{E}[O]_i + b$,
$y_i$ is the label of the $i$-th sample, and $|D_{val}|$ is the number of validation samples.
The positive weight vector $w$, a bias $b$, and a scaling factor $k$ are optimized by bounded minimization
of the log-loss using the L-BFGS-B algorithm \cite{zhu_algorithm_1997}.
Since the norm of the weight vector $w$ does not influence the objective function in Eq. \ref{eq:7b},
the unnormalized $w$ obtained from the optimization is rescaled to unit norm for the encoding in the control register.
By running the circuit with learned $w$ as detailed in Sec. \ref{sec:test-time}, Eq. \ref{eq:7} is estimated,
and the prediction for an input $x$ is then obtained as follows:
\begin{equation}\label{eq:8}
    \ytest(\xtest) = sign\left(\sigma(z(x)) -\frac{1}{2}\right) = sign(\mathbb{E}[O]+b/k).
\end{equation}

Considering a binary classification task, 
this formulation effectively performs a stacking of this logistic regression model on top of the internal classifiers, 
which act as a feature extractor layer. Considering the quantum classification algorithms 
and respective output probability formulations Eqs. \ref{eq:cos_p}, \ref{eq:dist_p}, \ref{eq:swap_p}, 
some of these methods require unit norm feature vectors for their prediction functions. 
However, in the data selection procedure, if we select a subset of features, 
the unit norm prerequisite is no longer valid.
This is represented by the $\delta_{c,i}$ coefficients in Eqs. \ref{eq:5-a} and \ref{eq:5-b}.
Despite this, even if classifiers are not used as intended, 
utilizing a weight optimization procedure, such as the one we considered, 
allows the suboptimal internal classifiers' outputs ($\beta$) to serve as informative features leveraged through logistic regression.

\section{Experiments}\label{sec:exp}
In this section we report empirical results of the performance of the proposed weighted ensemble.

\subsection{Methodology}
We implemented the scheme in Python with Qiskit \cite{ibm_qiskit_2024, javadi-abhari_quantum_2024}, 
and tested it on 11 real-world datasets from the UCI Machine Learning Library \cite{kelly_home_2024}, 
pre-processed to be suitable for a binary classification task. 
The implementation is available at \url{https://github.com/emiliantolo/quantum-ensemble-method}. 
The datasets are reported in Table \ref{tab:datasets}. 
\begin{table}[b]
    \centering
    \begin{tabular}{c c c c} \hline
        \textbf{Name} & \textbf{\# samples} & \textbf{\# features} & \textbf{Class balance} \\ \hline
        iris\_setosa\_versicolor       & 100 & 4  & balanced (50/50)        \\
        iris\_setosa\_virginica        & 100 & 4  & balanced (50/50)        \\
        iris\_versicolor\_virginica    & 100 & 4  & balanced (50/50)        \\
        vertebral\_column\_2C          & 310 & 6  & unbalanced (100/210)    \\
        seeds\_1\_2                    & 140 & 7  & balanced (70/70)        \\
        ecoli\_cp\_im                  & 220 & 7  & unbalanced (77/143)     \\
        glasses\_1\_2                  & 80  & 9  & almost balanced (42/38) \\
        breast\_tissue\_adi\_fadmasgla & 71  & 9  & unbalanced (49/22)      \\
        breast\_cancer                 & 80  & 9  & almost balanced (44/36) \\
        accent\_recognition\_uk\_us    & 80  & 12 & unbalanced (63/17)      \\
        leaf\_11\_9                    & 30  & 14 & almost balanced (14/16) \\ \hline
    \end{tabular}
    \caption{Considered datasets.}
    \label{tab:datasets}
\end{table}
The results have been computed with a Monte Carlo cross-validation technique \cite{xu_monte_2001}, 
with a 80\%--20\% training--validation dataset split and 10 independent runs. 
We employed the data with three different data normalization techniques. 
Specifically \textit{none} refers to data as it is. Instead, \textit{std} standardizes to mean 0 and standard deviation 1, 
then considering a training set $\{x_1, ..., x_n\}$ the $j$-th feature of the $i$-th training instance is defined as
\begin{equation}
    \operatorname{std}(\xtrain_i^j) = \frac{\xtrain_i^j - \mu({x_1^j, ..., x_n^j})}{\sigma({x_1^j, ..., x_n^j})},
\end{equation}
where $\mu({x_1^j, ..., x_n^j})$ and $\sigma({x_1^j, ..., x_n^j})$ are mean and standard deviation of the $j$-th feature of the training set. 
With \textit{minmax} each feature belongs to $[0, 1]$, then the normalization is defined as
\begin{equation}
    \operatorname{minmax}(\xtrain_i^j) = \frac{\xtrain_i^j - \min_{i=1, ..., n} (\xtrain_i^j)}{\max_{i=1, ..., n} (\xtrain_i^j) - \min_{i=1, ..., n} (\xtrain_i^j)},
\end{equation}
where $\min_{i=1, ..., n} (\xtrain_i^j)$ and $\max_{i=1, ..., n} (\xtrain_i^j)$ are the minimum and maximum features of the $j$-th feature in the training set. 
The normalization parameters have been computed on the training set samples for each dataset split. For the \textit{minmax} normalization, 
test points exceeding the $[0, 1]$ bound are clipped.

Regarding the execution, \textit{statevector simulation} refers to exact results computed based on the statevector, 
while \textit{local simulation} indicates circuit execution in a noiseless environment with 8192 repetitions using the Aer simulator.
It is worth noting that the number of shots includes both successful and unsuccessful data selections,
hence the effective number of shots used for the estimation of the outputs is lower.
The weighted ensemble adopts the data selection scheme described above and reported in Algorithm \ref{alg:selection}, 
thus the selection is performed by measuring the ancilla in 0 after the action of the first index and feature qubits, 
and the aggregation is by means of logistic regression which parameter optimization is performed 
using the SciPy package \cite{virtanen_scipy_2020}. We considered positive weights, 
constraining the optimizer with a positive bound. 
Moreover, we utilized the training set to train the weights at train-time.

\subsection{Results}

\begin{figure}[t!]
    \centering
    \includegraphics[width=1\linewidth]{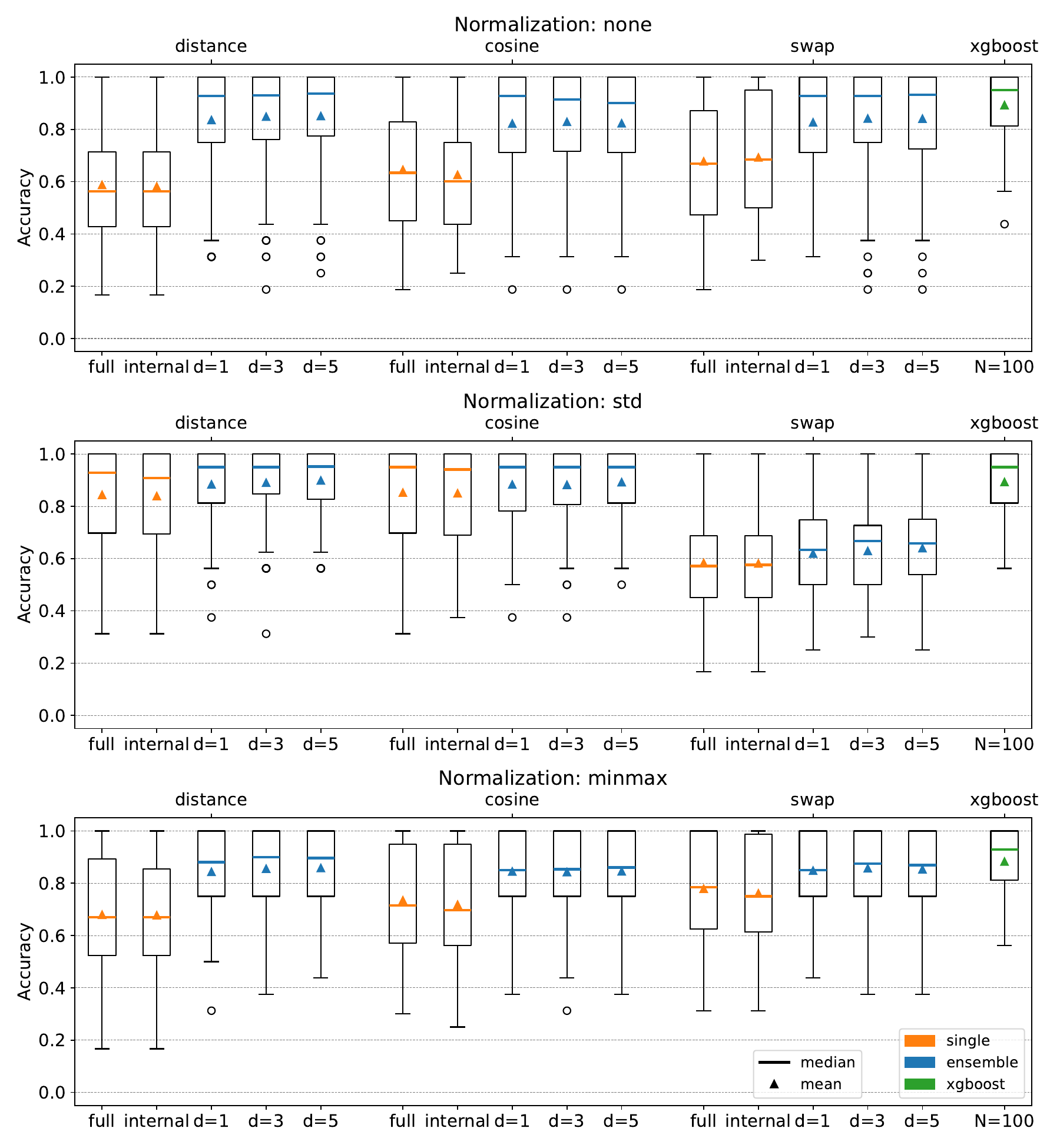}
    \caption{Accuracy performance with \textit{statevector simulation} of single classifiers (full), 
    ensemble's internal classifiers (internal), weighted ensemble and XGBoost in terms of aggregated results 
    over 10 Monte Carlo runs for 11 real-world datasets. 
    The considered internal models are the quantum cosine, quantum distance and SWAP test classifiers. 
    Multiple control register sizes (d) and normalization techniques are taken into account.}
    \label{fig:res}
\end{figure}

We considered the classifiers described above, namely the quantum cosine classifier, 
quantum distance classifier, quantum SWAP test classifier. 
We tested the single quantum classifiers using the full training set and simulated 
the execution of the ensemble's internal classifiers by applying the data selection procedure 
and feeding the resulting subset of the original training set to the classifier. 
We evaluated ensembles with various sizes of the control register (where $d$ indicates the number of control qubits).

Considering \textit{statevector simulation}, we reported the results in Figure \ref{fig:res}. 
In the ideal case, we observed a performance degradation of the internal classifier compared to the single classifier on average, 
with some combinations of classifier and data normalization being more affected by data selection. 
Despite this, the weighted ensemble seems to effectively utilize these weaker internal classifiers, 
showing a noticeable average accuracy improvement across all considered data normalization methods and classifiers. 
Moreover, the mean accuracy improves slightly on average with increasing control register size. 
Additionally, we included XGBoost \cite{chen_xgboost_2016}, a parallel implementation of boosted decision trees, 
as a performance reference for classical ensembles. While it typically delivers the best performance, 
it is matched in accuracy by the weighted ensemble in certain conditions with these datasets.

Regarding the quantum cosine and distance classifiers, they exhibit similar behavior, 
highlighting their relationship, particularly considering their close prediction functions. 
This similarity is also reflected in their performance when used as feature extractors within the ensemble, 
where they show comparable performance across various data normalizations. 
Specifically, the best classification performance is achieved with \textit{std} standardization. 
Conversely, \textit{minmax} normalization results in some information loss. 
While it starts with acceptable accuracy, 
the ensemble's performance ultimately falls below that of non-normalized data (\textit{none}).

Similarly, for the quantum SWAP test classifier, the ensemble with \textit{none} achieves the best average performance, 
even though the single classifier performs best with \textit{minmax}. 
On the other hand, \textit{std} normalization, which centers features in zero, 
deteriorates the performance of the SWAP test classifiers since it is based on state fidelity, 
where squaring the values causes the sign to disappear.

\begin{figure}[t!]
    \centering
    \includegraphics[width=1\linewidth]{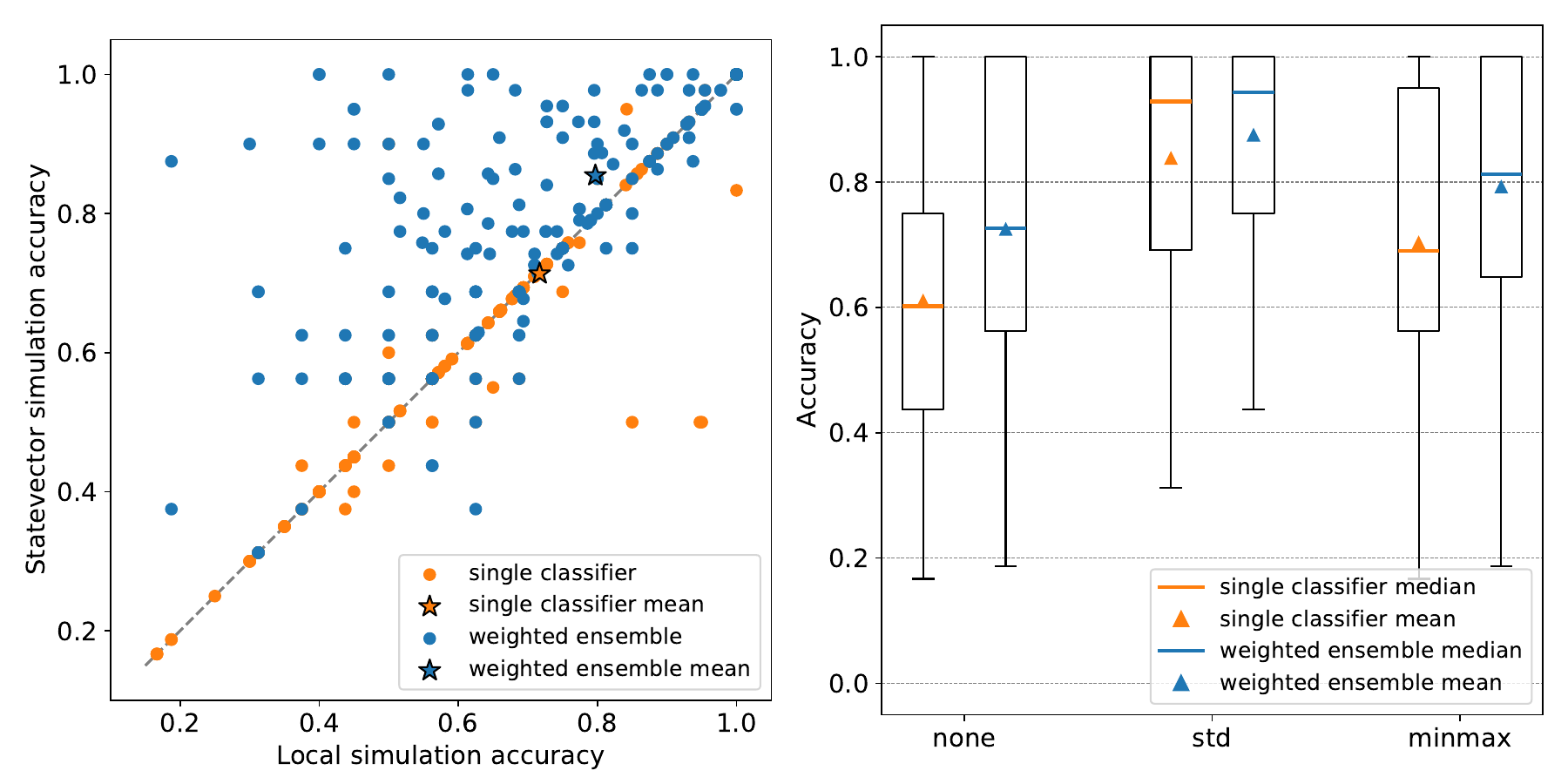}
    \caption{Accuracy performance of \textit{local simulation} with 8192 shots, 
    considering the quantum distance classifier on 11 datasets over 10 Monte Carlo runs. 
    The left panel shows the accuracy comparison between \textit{local simulation} and \textit{statevector simulation}, 
    where points above the bisector indicate accuracy degradation. 
    The right panel presents the accuracy comparison of single and ensemble classifier ($d=5$) with \textit{local simulation} 
    for the different normalization techniques.}
    \label{fig:plot}
\end{figure}

We conducted reduced-set testing using \textit{local simulation} to demonstrate the impact of circuit repetitions. 
Due to computational constraints, we employed only the quantum distance classifier with the same data as the previous tests. 
The single distance classifier shows a minimal decrease in average accuracy with shot noise, 
as fluctuations of mispredicted points near the prediction threshold balance out those of correctly predicted points, 
implying the inaccuracy of the single classifier near the threshold. 
On the other hand, the weighted ensemble is more affected by measurement repetitions, 
likely due to error propagation in the optimization process. 
However, it still maintains an accuracy advantage over the single classifier. 
These results are displayed in Figure \ref{fig:plot}.

\section{Discussion}\label{sec:disc}
The proposed weighted quantum ensemble method demonstrates a significant performance improvement over individual quantum classifiers. 
This method allows for a theoretical exponential ensemble size relative to the control register dimension. 

In terms of computational complexity, 
the proposed weighted ensemble introduces moderate overhead compared to a single classifier. 
At test time, the additional steps consist of preparing the control register of size $d$
(action of $U_w$ in Figure \ref{fig:circuit-test})
and executing the data selection ($U_p$ and MCX in Figure \ref{fig:circuit-test}),
both of which can be implemented with shallow circuits in practice.
Since $U_w$ is independent of the classifier data encoding $U_d$,
they can be executed in parallel prior to the data selection.
Moreover, the data selection circuit $S$, detailed in Algorithm \ref{alg:selection}, 
scales polynomially with the number of control and data qubits ($T_S \sim \operatorname{poly}(n,m,d)$).
The classifier execution $C$ remains unchanged.
Specifically, if $T_{QC} = O(N_{shots} \cdot (T_{U_d} + T_C))$ is the time complexity of the single quantum classifier,
for $N_{shots}$ repetitions of the circuit, then the complexity of the weighted ensemble at inference time is 
\begin{equation}\label{eq:complexity-test}
    T_{QE} = O\left(N_{shots} \cdot \frac{1}{F} \cdot (\operatorname{max}(T_{U_d},T_{U_w}) + T_S + T_C)\right),
\end{equation}
where $F<1$ is the fraction of the training set selected by the data selection circuit.
The term $F$ reflects the fact that, on average, to obtain a useful computational path,
the circuit must be executed $1/F$ times to select the correct data subset.
In practice, $T_{U_w}$ can often be neglected since it can be in parallel to $U_d$, 
and usually smaller due to a small number of control qubits $d$.
Thus, the ensemble's inference complexity is comparable to that of the single classifier.

At training time, the ensemble circuit is executed to collect the outputs of the $2^d$ internal classifiers for a validation set of size $|D_{val}|$.
The time complexity for the training circuit execution is 
\begin{equation}\label{eq:complexity-train}
    T_{QE_\text{train}} = O\left( 2^d \cdot |D_{val}| \cdot N_{shots} \cdot \frac{1}{F} \cdot (T_{U_d} + T_S + T_C)\right),
\end{equation}
where $N_{shots} \cdot 1/F$ is the effective number of measurements for each internal classifier and validation sample.
Notably, at this stage, there is no need to encode weights in the control register.
The training procedure is hybrid as it requires an arbitrary classical optimization procedure to compute the weights $w$.
Since the optimization is performed on the outputs of the internal classifiers,
the complexity of the optimization procedure is at least $O(2^d \cdot |D_{val}|)$.
However, this is a one-time, offline procedure.

That said, the primary focus is not on computational complexity but rather on the performance enhancement 
achievable with the proposed scheme, which is evident even with limited control register sizes.
Empirical results showed that the proposed weighted ensemble outperformed the single quantum classifier consistently,
across various datasets and data normalization methods. 
This indicated that the ensemble successfully utilized the diversity of the weaker internal classifiers,
introduced by the data selection. Moreover, the ensemble showed an increased robustness to data normalization,
with instances where the ensemble with non-standardized data outperformed the ensemble with the standardization
that showed the best result with the single classifier. 

In the local simulation setting, where shot noise is present, 
the weighted ensemble was more sensitive to noise than the single classifier. 
This sensitivity is likely due to the optimization stage of the ensemble, 
which can be negatively affected by inaccuracies in individual outputs.
Nonetheless, the ensemble showed acceptable resilience to noise, 
as it outperformed the accuracy of the single classifier.

Indeed, for linear classifiers, such as distance and cosine quantum classifiers, 
the weighted aggregation can be interpreted as a modified dataset with multiplicative factors on training points and features. 
This allows for an alternative execution of the weighted ensemble at test-time by using a single quantum classifier 
on the modified dataset, avoiding the data selection circuit logic 
but requiring a change in the data encoding procedure of the initial state.

For non-linear classifiers, such as those using the SWAP test, 
the weighted ensemble can be interpreted as a two-layer neural network (NN) with some missing connections in the first layer 
and the second layer's weights trained by an optimization procedure. 
A similar NN structure based on the SWAP test has been proposed in \cite{pastorello_scalable_2024}, 
which considers a classical combination of quantum modules, 
whereas the proposed method has a quantum aggregation of SWAP test based classifiers.
To assess the non-linear prediction capabilities and correctness of our approach, 
we tested it on a synthetic 2-dimensional XOR dataset. As expected, the linear classifiers and related ensembles struggled, 
while the ensemble using the non-linear SWAP test achieved nearly perfect accuracy.

Additionally, a majority voting aggregation strategy can be implemented 
by sampling the circuit defined for the training procedure. 
This allows the individual outputs of the internal classifiers to be estimated and then combined 
in a bagging fashion via majority voting. 
Sampling with replacement-like data sampling may be advantageous 
and can be simulated by using a larger training dataset, 
employing multiple copies of the original training data, and allowing for training point repetitions 
in each subset up to the number of original training set copies.

\section{Conclusion}\label{sec:concl}
In this work, we proposed a weighted homogeneous quantum ensemble method 
that leverages instance-based quantum classifiers to improve classification performance. 
The method introduces diversity through training point and feature subsampling in superposition, 
allowing for a quantum-parallel execution of the diverse internal classifiers, 
which outputs are combined with weights. 
The weight learning procedure is hybrid, 
where the quantum circuit is executed to collect the outputs of the internal classifiers, 
and the weights are trained classically. 
The deployment is quantum, encoding the weights in the circuit at test-time.

The empirical evaluation on real-world datasets demonstrated the effectiveness of the proposed method. 
The results showed improved accuracy over the single quantum classifier, 
both in the exact case by considering results computed with expectation values, 
and with shot noise considering circuit measurement repetitions. 
We compared the accuracy of the quantum ensemble with XGBoost and showed comparable performance 
with some internal classifiers and data normalizations.

Moreover, let us remark that, while variational quantum circuits have become central in current QML research, non-variational approaches like the proposed one based on ensembles of similarity-based classifiers, avoiding gradient-based optimization for circuit training and related issues (e.g. barren plateaus), align naturally with the NISQ constraints. In this way, in our opinion, quantum ensembles of quantum (non-variational) classifiers represent an advantageous solution in QML. 

Future directions include exploring different diversity introduction mechanisms not based on data subsetting, 
such as parametric controlled rotation operations. 
Overall, the proposed quantum weighted ensemble method offers a robust and flexible framework 
for improving classification accuracy in quantum machine learning.

\section*{Acknowledgements}
\noindent
E.T. was supported by the MUR National Recovery and Resilience Plan (PNRR) M4C1I4.1, 
funded by the European Union under NextGenerationEU. 
Views and opinions expressed are however those of the author(s) only 
and do not necessarily reflect those of the European Union or The European Research Executive Agency. 
Neither the European Union nor the granting authority can be held responsible for them.
D.P. was supported by project SERICS (PE00000014) under the MUR National Recovery and Resilience Plan funded by the European Union -- NextGenerationEU. D.P. is a member of the ``Gruppo Nazionale per la Fisica Matematica (GNFM)'' of the ``Istituto Nazionale di Alta Matematica ``Francesco Severi'' (INdAM)''

\setcounter{footnote}{0}
\renewcommand{\thefootnote}{\alph{footnote}}

\printbibliography

\end{document}